\documentstyle[prl,preprint,aps]{revtex}
\begin{document}
\draft
\title{ARPES study of the superconducting gap and pseudogap in 
$Bi_2Sr_2CaCu_2O_{8+x}$}
\author{
        H. Ding,$^{1,2}$
        J. C. Campuzano,$^{1,2}$
        M. R. Norman,$^1$
        M. Randeria,$^3$ 
        T. Yokoya,$^4$
        T. Takahashi,$^4$
        T. Takeuchi,$^{5}$        
        T. Mochiku,$^6$
        K. Kadowaki,$^7$
        P. Guptasarma,$^1$
        and D. G. Hinks$^1$
       }
\address{
         (1) Materials Sciences Division, Argonne National Laboratory,
             Argonne, IL 60439 \\
         (2) Department of Physics, University of Illinois at Chicago,
             Chicago, IL 60607\\
         (3) Tata Institute of Fundamental Research, Mumbai 400005,
India\\
         (4) Department of Physics, Tohoku University, 980 Sendai, Japan\\
         (5) Department of Crystalline Materials Science, Nagoya 
             University, Nagoya 464-01, Japan\\
         (6) National Research Institute for Metals, Sengen, Tsukuba,
             Ibaraki 305, Japan\\
         (7) Institute of Materials Science, University of Tsukuba, 
             Ibaraki 305, Japan\\
         }
\maketitle
\begin{abstract}
In this paper, we review some of our ARPES results on the 
superconducting and pseudo gaps in $Bi_2Sr_2CaCu_2O_{8+x}$. We 
find that optimally and overdoped samples exhibit a d-wave
gap, which closes at the same temperature, $T_c$, for all {\bf k} points. 
In underdoped samples, a leading edge gap is found up to a temperature  
$T^{*}>T_c$. We find that $T^{*}$ scales with the maximum low 
temperature gap, increasing as the doping is reduced. The 
momentum dependence of the pseudogap is similar to that of the 
superconducting gap; however, the pseudogap closes at different
temperatures 
for different {\bf k} points.
\end{abstract}
The nature of the energy gap has been an important issue in the field of
high 
temperature superconductivity. In conventional BCS superconductors, 
there is an isotropic energy gap (s-wave order parameter) below the 
critical temperature, $T_c$, which is a direct consequnce of electron 
pairing mediated by phonons. However, high $T_c$ superconductors 
appear to be very different in this respect. First, the 
superconducting gap is highly anisotropic. In fact, 
an intense debate on the s- or d-wave symmetry of the order parameter 
has dominated this field for the past several years \cite{LEGGETT}. 
More recently, attention has focused on the pseudogap which is
formed above $T_c$ in underdoped samples\cite{LEVI}.
The origin of the pseudogap and its relation to the 
superconducting gap below $T_c$ is still a subject of great 
controversy. Angle-resolved photoemission spectroscopy (ARPES) has 
played a major role in understanding the superconducting and pseudo   
gaps due to its much improved energy resolution and unique 
momentum-resolving capability. ARPES has been particularly successful 
in measuring $Bi_2Sr_2CaCu_2O_{8+x}$ (Bi2212) single crystals, because 
when these materials are cleaved to expose a clean surface, the potential 
at the surface is not significantly altered due to the van der Waals 
bonds between the two BiO cleavage planes. Furthermore, the electronic 
structure of Bi2212 is nearly perfectly two-dimensional, which 
simplifies the interpretation of the data. 

In this paper we review our ARPES results on the superconducting and
pseudo 
gap, and evidence for a connection between them. In summary, we have found 
that, in optimally and overdoped samples of Bi2212, a d-wave
superconducting 
gap closes at $T_c$ for all {\bf k} points on the Fermi surface. However,
in 
underdoped samples the superconducting gap below $T_c$ smoothly evolves
into 
a pseudogap above $T_c$, which closes at different temperatures for
different 
{\bf k} points on the Fermi surface.   

Earlier ARPES results had found the superconducting gap in optimally doped 
Bi2212 to be highly anisotropic\cite{SHEN,PRL95}. However, due to either 
sparse sampling of {\bf k} points or the complications caused by 
the superlattice in the BiO layers, neither study conclusively established
the 
momentum dependence of the excitation gap. We later carried out more
careful
measurements in a number of near-optimally doped Bi2212 samples, 
using a dense sampling of {\bf k} points and avoiding superlattice 
bands\cite{GAP96}. A clear d-wave picture emerged from those 
experiments. As an example, in Fig.~1 we plot fitted gap values of a 
lightly overdoped sample ($T_c$ = 87K) at different {\bf k} points along 
the CuO Fermi surface. Great care was taken in identifying Fermi 
surface crossing points. Because of a narrow (resolution-limited) spectral 
lineshape in the superconducting state, a simple BCS spectral function 
broadened by (energy and momentum) resolution can be used to fit the 
spectra. Although ARPES does not measure the order parameter, combining
the 
observation that the excitation gap in Fig.~1 follows a simple d-wave
form, 
$|\cos(k_x)-\cos(k_y)|$, with phase sensitive experiments\cite{TSUEI}, 
one can safely conclude that the order parameter in Bi2212 has a 
nearly pure $d_{x^2-y^2}$ form.  

We now discuss the temperature dependence of the gap. It is generally 
difficult to extract the temperature dependence of $\Delta$ 
because the spectral peaks acquire significant widths at higher 
temperatures, and the assumption of negligible linewidths for fitting 
the spectra is no longer valid. As an example, in Fig.~2 
we show the full width at half-maximum (FWHM) of the spectral peak near
the 
$(\pi,0)$ point as a function of 
temperature. We note that this width does not provide the inverse
lifetime 
of the state, because at low temperatures the width is given by the 
experimental resolution, while at higher temperatures (above $T_c$) 
the Fermi function controls the leading edge width. Nevertheless, the plot
provides 
an indication 
of the very steep decrease in lifetime on approaching $T_c$. We can  
obtain a crude estimate of the temperature dependence of the gap by
plotting 
the midpoint of the leading edge of the spectra as a function of
temperature, 
as shown in Fig.~3. Again, the sample is a lightly overdoped 87K one. It 
is clear from Fig.~3 that gaps at different {\bf k}-points vanish at the 
same temperature, close to the bulk $T_c$. This result is significant, 
because it indicates that in lightly overdoped samples, there is 
only one gap at the Fermi surface, the superconducting gap, consistent 
with the fact that the momentum dependence of this gap follows a simple
d-wave function.

The picture changes dramatically in underdoped samples. ARPES experiments 
have shown that there is a leading edge gap above
$T_c$\cite{LOESER,NATURE}. 
This can be clearly seen from Fig.~4, where we plot spectra at the 
$(\pi,0) - (\pi, \pi)$ Fermi surface
point of an underdoped ($T_c$ = 83K) sample at six different temperatures. 
Note that above $T_c$, i.e. at 90K, there is a sizeable (16 meV) shift 
between the leading edge of the sample (solid line) and that of
polycrystalline 
Pt (dotted line) which is used as a chemical potential reference. This 
pseudogap eventually disappears at a much higher temperature $T^*$
($\approx 200K$ in this case).

It is significant to note that there are always two features in the 
spectra, one that is related to the quasiparticle peak in the 
superconducting state, and gives rise to the sharp leading edge in the 
pseudogap state, and another feature at higher binding energy, described 
in the literature as the ``hump''\cite{DESSAU}.  The pseudogap that we
describe 
here is associated with the feature at low binding energy, the leading 
edge gap\cite{UDFS}.

It has been found that $T^*$ increases with deceasing doping in the 
underdoped region, and merges with $T_c$ in the overdoped 
region\cite{NATURE}, as shown in Fig.~5. In  Fig.~5 we 
also plot the position of the sharp coherent peak near 
$(\pi,0)$ (see first panel of Fig.~4) as a function of doping, or carrier 
concentration x. Since this sharp peak is essentially resolution limited, 
one can regard the position of its maximum as the value of the gap, 
$\Delta(0)$. Despite some considerable sample-to-sample 
variation, $\Delta(0)$ follows the general trend of increasing with 
decreasing x. In fact, $\Delta(0)$ seems to scale with 
$T^*$, not with $T_c$. This is consistent with 
theories which predict that $T_c$ is controlled 
by a phase stiffness temperature\cite{RANDERIA,EMERY}, and not by the 
temperature at which a pairing gap opens. On the 
other hand, one may argue that the gap near $(\pi,0)$ is no longer the 
superconducting gap, since it has no relationship to $T_c$.

Let us address this problem by looking at some experimental evidence. 
Temperature dependent measurements in underdoped samples, shown 
in Fig.~6, reveal a gap that smoothly evolves through $T_c$, 
suggesting that the gaps below and above $T_c$ have the same origin,
i.e the pseudogap is closely related to the superconducting 
gap.  We have also found that the low temperature (T = 14K) gap of an 
underdoped ($T_c$ = 83K) sample has a very similar momentum dependence 
as the gap of the overdoped 87K sample which has the d-wave gap shown in 
Fig.~1\cite{NATURE}. This is a strong indication that in underdoped 
samples the gap below $T_c$ near $(\pi,0)$ is still the 
superconducting gap. 
It is interesting to note that, although having little 
effect on the the gap size near $(\pi,0)$, $T_c$ has a strong effect 
on the lineshape.

As mentioned above, in optimally and overdoped samples
the superconducting gap closes at $T_c$ for all {\bf k} points. 
What will be the case in underdoped samples? To answer this question, 
we have recently performed ARPES measurements on several underdoped 
Bi2212 samples. To our surprise, we found that pseudogaps at different 
{\bf k} points close at different temperatures, in marked contrast 
with the result obtained in optimally doped samples\cite{DESTRUCTION}.
Fig. 7 shows one example where we plot midpoint shifts for an 85K 
underdoped sample at three {\bf k} points. Point (a) is near the 
$(\pi,0)$ to $(\pi,\pi)$ crossing, with points (b) and (c) progressively 
closer to the node direction ($\Gamma - Y$), as shown in the inset.
>From this plot, we find that the pseudogap closes at point (a) at a 
temperature above 180 K, at point (b) at 120 K, and at point (c) 
just below 95 K. If we view these data as a function 
of decreasing temperature, we see that the pseudogap first 
opens near $(\pi,0)$ and progressively gaps out larger and larger 
portions of the Fermi surface, leading to gapless arcs shrinking 
with decreasing temperature, eventually collapsing to the point 
nodes of the $d_{x^2-y^2}$ superconducting ground state below $T_c$.

In conclusion, we have found that in optimally and overdoped samples, 
a d-wave superconducting gap closes at the same temperature, $T_c$, 
for all {\bf k} points. However, in underdoped samples, the
superconducting 
gap below $T_c$ smoothly evolves into a pseudogap above $T_c$, 
which closes at different temperatures for different {\bf k} points. 
This suggests an intimate, but non-trivial, relation between the 
superconducting gap and the pseudogap.

This work was supported by the National Science Foundation 
DMR 9624048, and DMR 91-20000 through the Science and Technology 
Center for Superconductivity, and the U. S. Dept. of Energy,
Basic Energy Sciences, under contract W-31-109-ENG-38. The 
Synchrotron Radiation Center is supported by the NSF.

\begin{figure}
\caption{The superconducting gap of an 87K overdoped Bi2212, 
extracted from fits, 
versus angle on the Fermi surface (filled circles) compare to
a d-wave gap (solid curve).
Locations of measured points and the Fermi surface are shown in the inset.
}
\label{fig1}
\end{figure}

\begin{figure}
\caption{FWHM of the spectral peak as a function of
temperature in a slightly 
overdoped 87K Bi2212 sample. 
}
\label{fig2}
\end{figure}

\begin{figure}
\caption{The superconducting gap of an 87K overdoped Bi2212 
(estimated by midpoints of the 
leading edge of the spectra) at two different ${\bf k}$ points
(indicated in the 
inset) as a function of temperature. Note both gaps close near $T_c$. 
}
\label{fig3}
\end{figure}

\begin{figure}
\caption{ARPES spectra at the Fermi surface along the $M - Y$ direction  
for an 83K underdoped Bi2212 sample at various temperatures (solid
curves).
The dotted curves are reference spectra from polycrystalline Pt.
}
\label{fig4}
\end{figure}

\begin{figure}
\caption{Low temperature superconducting gap $\Delta(0)$ near
$(\pi, 0)$ measured by peak 
positions (circles), $T^*$ (triangles for determined 
values, squares for lower bounds), and $T_c$ (dashed line) 
as a function of carrier concentration, x. Note a similar trend for both 
$\Delta(0)$ and $T^*$.}
\label{fig5}
\end{figure}

\begin{figure}
\caption{Midpoints of the leading edge of the spectra for an 83K
underdoped 
Bi2212 near $(\pi,0)$ as a function of temperature. Note the smooth 
evolution through $T_c$.
}
\label{fig6}
\end{figure}

\begin{figure}
\caption{Midpoints of the leading edge of the spectra for an 85K
underdoped 
Bi2212 at three {\bf k} points 
(indicated in the Brillouin zone) as a function of temperature. Note
the closing of the spectral gap at different temperatures for different
{\bf k}.
}
\label{fig7}
\end{figure}

\end{document}